\documentclass[fleqn,10pt]{wlscirep}
\usepackage{amssymb, amsmath, amsfonts, amsthm, graphicx, subfig, color,soul,array,booktabs,lineno,prettyref,cite,hyperref}

\title{Quantification of Healthy Red Blood Cell Removal and Preferential Invasion of Reticulocytes in \textit{Macaca mulatta} during \textit{Plasmodium cynomolgi} Infection}
\begin{document}
\author[1]{Yi H. Yan}
\author[3]{Jacob B. Aguilar}
\author[1]{Elizabeth D. Trippe}
\author[1,3,*]{Juan B. Gutierrez}
%\author{The MaHPIC Consortium}

\affil[1]{University of Georgia, Institute of Bioinformatics, Athens, 30602, USA}
\affil[2]{Universidad del Quindio (Colombia)}
\affil[3]{University of Georgia, Department of Mathematics, Athens, 30602, USA}

\affil[*]{Corresponding author:jgutierr@uga.edu}
%\affil[+]{These Authors Contributed To This Work Equally}
\begin{abstract}

We derived an ordinary differential equation model to capture the disease dynamics during blood-stage malaria. The model was directly derived from an earlier age-structured partial differential equation model. The original model was simplified due to experimental constraints. Here we calibrated the simplified model with experimental data using a multiple objective genetic algorithm. Through the calibration process, we quantified the removal of healthy red blood cells and the the preferential infection of reticulocytes during \textit{Plamodium cynomolgi} infection of \textit{Macaca mulatta}. The calibration of our model also revealed the existence of host erythropoietic response prior to blood stage infection. 

\end{abstract}
\date{\today}

\maketitle 
%%%%%%%%%%%%%%%%%%%%%%%%%%%%%%%%%%%%%%%%%%%%%%%%%%%%%%%
\section{Introduction}
%%%%%%%%%%%%%%%%%%%%%%%%%%%%%%%%%%%%%%%%%%%%%%%%%%%%%%%
The goal of this article is to quantify the removal of healthy red blood cells and preferential invasion of reticulocytes during malaria infection, and find their associations with molecular phenomena through a modeling approach. The disease malaria is caused by \textit{Plasmodium} parasites. Out of the five human \textit{Plasmodium} species capable of causing malaria, \textit{Plasmodium falciparum} and \textit{Plasmodium vivax} account for the majority of human malaria infections. \textit{P. falciparum} is responsible for the majority of malaria-related mortality and is most prevalent in sub-Saharan Africa \cite{world2016world}. 

The \textit{Plasmodium} life cycle is comprised of several stages. The infection process in humans starts with the injection of sporozoites by mosquitoes into the skin of the host. This is followed by the liver stage, during which the inoculated sporozoites grow and multiply asexually within hepatocytes for 1-2 weeks to produce merozoites. The newly produced merozoites emerge from the liver and enter the blood stream. The blood-stage infection starts immediately after the hepatic stage. The parasite's blood-stage infection in both human and non-human primates generally has a regular cycle of 24 to 72 hours depending on the species of the \textit{Plasmodium} parasite \cite{bozdech2003transcriptome, collins2001plasmodium}. The parasites invade healthy red blood cells (RBCs) and replicate asexually, remodeling and ultimately destroying the RBCs in the process. The destruction of RBCs during blood-stage malaria infection sometimes results in severe anemia, which is one of the major complications of malaria and a leading cause of mortality.

During blood-stage malaria infection, the majority of red blood cell loss has been attributed to the removal of healthy red blood cells \cite{collins2003retrospective,jakeman1999anaemia,price2001factors}. The hemodynamic model described by Yan \textit{et al} \cite{yan2015mathematical} characterizes this phenomenon through a mechanistic model taking into account the interaction between healthy RBCs, infected RBCs, cells form the innate immune system, and cells from the adaptive immune system. This model captures the clinical outcomes of a malaria infection: resistance to the disease, susceptibility, and resilience (chronic infection with mild symptoms). That model has been simplified to account for experimental constraints, and it has been calibrated in this article with the data of \textit{Macaca mulatta} infected with \textit{P. cynomolgi} described by Joyner \textit{et al} \cite{joyner2016plasmodium}. \textit{Plasmodium cynomolgi} is a malaria parasite that infects old world monkeys; it is physiologically and evolutionarily similar to \textit{P. vivax}  \cite{warren1966biology,waters1993evolutionary}. %  It exhibits 48-h erythrocytic cycle during blood stage infection and preferentially infects reticulocytes \cite{warren1966biology}. 

This paper is organized as follows:  Section 2 provides the description of the experimental data. Section 3 contains the detailed derivation and description of our hemodynamic model. Section 4 describes the model calibration process. Section 5 provides an overview of the calibration results. Section 6 integrates our modeling result with transcriptomic data collected during the experiment. Section 7 describes an adjusted model based on our calibration results. Section 8 offers some conclusions, and discusses the biological significance of our results. 

%%%%%%%%%%%%%%%%%%%%%%%%%%%%%%%%%%%%%%%%%%%%%%%%%%%%%%%
\section{Experimental Description}
%%%%%%%%%%%%%%%%%%%%%%%%%%%%%%%%%%%%%%%%%%%%%%%%%%%%%%%
\begin{figure}[!ht]
	\centering
	\includegraphics[width = 0.9\textwidth]{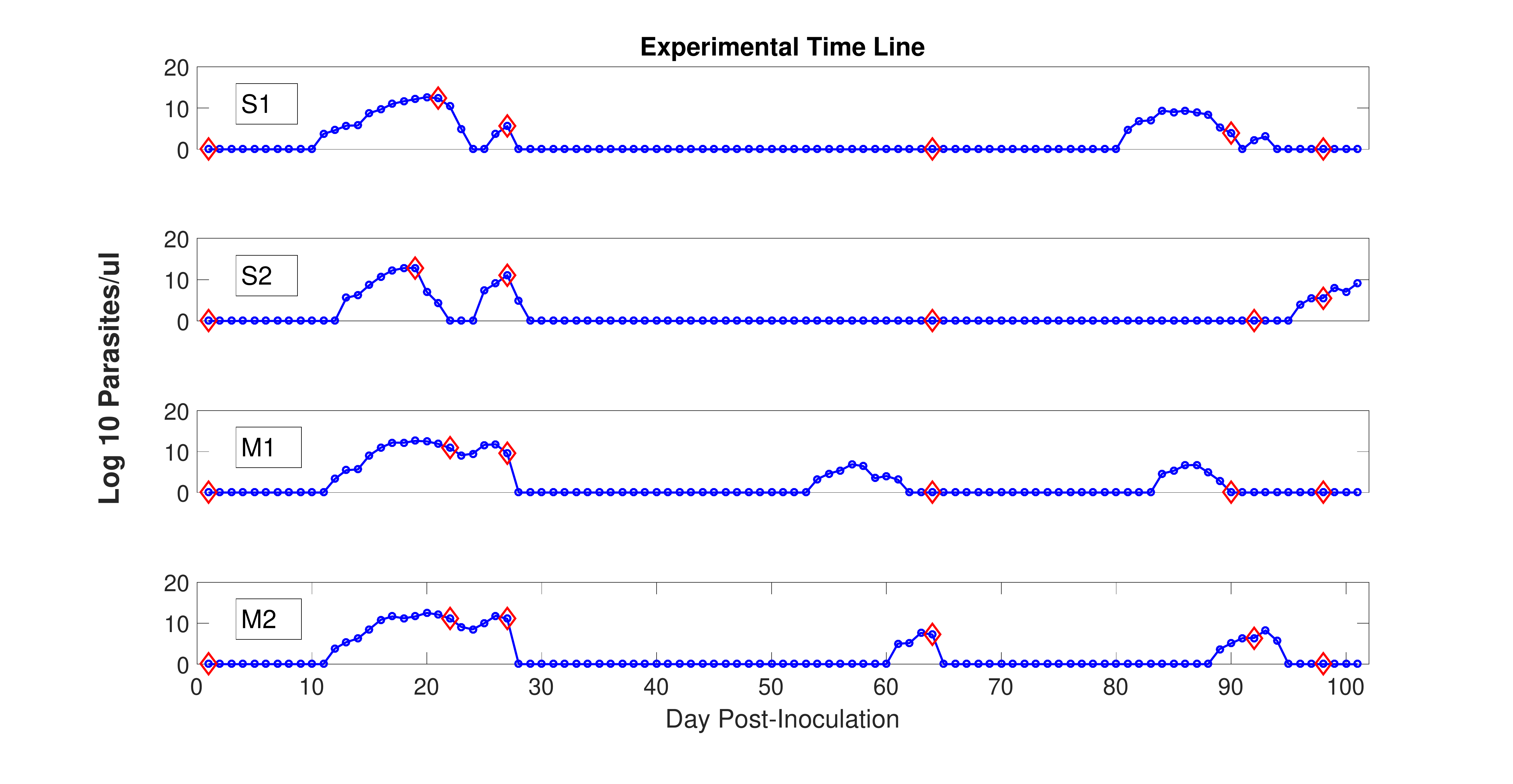}
	\caption[Experimental Setup]
    {Parasite count over the entire experiment. Transcriptome, cytokine, and immune profiling were conducted during the 6-time points marked with the red diamond. Subject S1 and S2 experienced severe malaria symptoms. Subject M1 and M2 experienced mild malaria symptoms.}
	\label{fig:Experimental Set Up2}
\end{figure}

The experimental design is shown in Fig \ref{fig:Experimental Set Up2}. The infected red blood cell, total red blood cell, and reticulocyte concentration were measured daily over the entire experiment. The four subjects, S1, S2, M1 and M2 are referred as Subject 1 $\sim$ 4 within this paper. Each red diamond corresponds to a specific time point where transcriptome, cytokine and immune profiling was conducted. Time point 2 (TP2) corresponds to the acute primary infection when parasite counts peaked. A detailed description of the experiment that collected this data can be found in Joyner \textit{et al.} \cite{joyner2016plasmodium}.

%%%%%%%%%%%%%%%%%%%%%%%%%%%%%%%%%%%%%%%%%%%%%%%%%%%%%%%
\section{Matching a Theoretical Model with Experimental Constraints}
%%%%%%%%%%%%%%%%%%%%%%%%%%%%%%%%%%%%%%%%%%%%%%%%%%%%%%%
The mechanistic partial differential equation model described by Yan \textit{et al.} \cite{yan2015mathematical}, has the  form presented below; the  detailed definition of variables, parameters and functions is presented in that paper, and also for completeness in the Appendix. 
\begin{equation} \label{eq:simplified system2}
	\begin{split}  
		\frac{\partial u}{\partial t}  +  \frac{\partial u}{\partial a} = &
			-\left(\sum_{i= 1}^T w_{i}(t)\theta_{i}+\sum_{i= 1}^Q s_{i}(t)\psi_{i} + h(a)\right) u(a, t) \\
			& - \gamma \kappa v(\alpha_{max}, t)p(u(a, t)), \\
		\frac{\partial v}{\partial t} + V\frac{\partial v}{\partial \alpha} = &-\left(\sum_{i= 1}^T w_{i}(t)\phi_{i} + \sum_{i= 1}^Q s_{i}(t)b_{i}(t)\right)v(\alpha, t), \\
		\frac{dw_{i}}{dt} = &o_{i}(t) - \beta_{i}w_{i}(t),\\ 
		\frac{ds_{i}}{dt} = &l_{i}(t) - \delta_{i}s_{i}(t),
	\end{split}
\end{equation}
subject to the following initial and boundary conditions:
%--------------------------------------
\begin{equation*}
	\begin{split}
		u(a, 0) = &g(a), \\
		u(0, t) = &f(t, \varphi(t)), \\ 
	  v(\alpha, 0) = &c(\alpha), \\
		v(0, t) = &\gamma \kappa \int_{a_{0}}^{a_{max}}v(\alpha_{max}, t)p(u(a, t))\, da. 
	\end{split}
\end{equation*}
With the experimental constraints, it was not possible to measure the delay in the erythropoietic response, and hence it was removed from this model. Thus the boundary condition $u(0, t) = x_4e^{x_5(T_0-T(t))}$, where $x_4$ is the baseline production of red blood cells and $x_5$ is a parameter controlling the speed of erythropoiesis response. $T_0$ is the sum of the steady-state concentration of reticulocytes and mature red blood cell and $T$ is the sum of the concentration of reticulocytes and mature red blood cell at time $t$. 

Under the explicit assumption, that the immune cell populations do not interact with the infected red blood cell population ($v$) and healthy red blood cell population ($u$) ($\phi_{i} = 0$, $\theta_{i} = 0$, $\psi_{i} = 0$, for all $i$ and $b_{i}(t) = 0$ for all $t$), the system reduces to: 
\begin{equation*} \label{eq:simplified system 2}
	\begin{split}  
		\frac{\partial u}{\partial t}  +  \frac{\partial u}{\partial a} = &
			-h(a)u(a, t) - \gamma \kappa v(\alpha_{max}, t)p(u(a, t)), \\
		\frac{\partial v}{\partial t} + V\frac{\partial v}{\partial \alpha} = &  0.
	\end{split}
\end{equation*}
%------------------------
Furthermore, under the explicit assumption that the initial age distribution of v is uniform, $v(\alpha, 0) = C$. Let 
%------------------------
\begin{equation}
	V(t)  = \int_0^{\alpha_{max}} v(\alpha,t)d\alpha, \nonumber
\end{equation}
%------------------------
where $V(t)$ denote the total amount of iRBC at time $t$. Then,
%------------------------
\begin{equation}
  V(0) = \alpha_{max}C. \nonumber
\end{equation}
%------------------------
Additionally, let $t_2 = t_1 + \epsilon, \epsilon > 0$, then: 
%------------------------
\begin{equation*}
  \begin{split}
  \int_0^{\alpha_{max}} v(\alpha,t_2)d\alpha = &\int_0^{\alpha_{max}} v(\alpha,t_1)d\alpha,\\&+\int_{t_1}^{t_2} \gamma \kappa \int_{a_{0}}^{a_{max}}v(\alpha_{max}, t)p(u(a, t))da dt,\\&-\int_{t_1}^{t_2} v(\alpha_{max}, t)dt.
  \end{split}
\end{equation*}
%------------------------
Because $\int_{a_{0}}^{a_{max}} p(u(a, t)) = 1$ for all $t$, the system further simplifies to: 
%------------------------
\begin{equation} \label{eq:VDeriv}
  \begin{split}
    \int_0^{\alpha_{max}} v(\alpha,t_2)d\alpha = &\int_0^{\alpha_{max}} v(\alpha,t_1)d\alpha,\\&+\int_{t_1}^{t_2} \gamma \kappa v(\alpha_{max}, t)dt,\\&-\int_{t_1}^{t_2} v(\alpha_{max}, t)dt.
  \end{split}
\end{equation}
%------------------------
Substituting $V$ into \prettyref{eq:VDeriv}, and rearranging the equation: 
%------------------------
\begin{equation}
  V(t_2) - V(t_1) = (\gamma \kappa - 1)\int_{t_1}^{t_2} v(\alpha_{max},t)dt. \nonumber
\end{equation}
%------------------------
Thus, 
%------------------------
\begin{equation}
  \int_{t_1}^{t_2} \dot{V} dt = (\gamma \kappa - 1)\int_{t_1}^{t_2}v(\alpha_{max},t)dt \nonumber
\end{equation}
%------------------------
and 
%------------------------
\begin{equation}\label{eq:VdotV1}
  \dot{V} = (\gamma \kappa - 1) \int_{a_{0}}^{a_{max}}v(\alpha_{max}, t)p(u(a, t))\, da.
\end{equation}
%------------------------
Because the age distribution of red blood cells are unknown throughout the experiment, and only reticulocyte and mature red blood cell populations are measured, we define $a_1$ as the precise time point where reticulocytes become mature red blood cells. We further define the total population of reticulocytes at time point $t$ as $R(t)$ and the total population of mature red blood cells at time point $t$ as $U(t)$. $R(t)$ and $U(t)$ have the following form:
%------------------------
\begin{equation}
  R(t) = \int_{0}^{a_1} u(a,t)da, \nonumber
\end{equation} 
%------------------------
and 
%------------------------
\begin{equation}
  U(t) = \int_{a_1}^{a_{max}} u(a,t)da, \nonumber
\end{equation}
%------------------------
thus 
%------------------------
\begin{equation}
  R(t) + U(t) = \int_{0}^{a_{max}} u(a,t)da. \nonumber
\end{equation}
%------------------------
Therefore, \prettyref{eq:VdotV1} can be rewritten as: 
%------------------------
\begin{equation}
  \dot{V} = (\gamma \kappa - 1)v(\alpha_{max}, t)\left(\frac{R}{R+U} + \frac{U}{R+U}\right). \nonumber
\end{equation}
%------------------------
Under the assumption that on average a small fixed percentage of infected red blood cells are bursting, then
%------------------------
\begin{equation}
	v(\alpha_{max}, t) = C_2V. \nonumber
\end{equation}
%------------------------
Which means: 
%------------------------
\begin{equation}
  \dot{V} = (\gamma \kappa - 1)C_2V\left(\frac{R}{R+U} + \frac{U}{R+U}\right). \nonumber
\end{equation}
%------------------------
Additionally, knowing that the infection rate of reticulocyte is different from mature red blood cells, we can modify the system to reflect this fact: 
%------------------------
\begin{equation}
	\dot{V} = (\gamma \kappa_1 - 1)C_2V\frac{R}{R+U} + (\gamma \kappa_2 - 1)C_2V\frac{U}{R+U}. \nonumber
\end{equation}
%------------------------
Let 
%------------------------
\begin{equation}
  x_3N = (\gamma \kappa_1 - 1)C_2, \nonumber
\end{equation}
%------------------------
and 
%------------------------
\begin{equation}
  x_6N = (\gamma \kappa_2 - 1)C_2. \nonumber
\end{equation}
%------------------------
Then 
%------------------------
\begin{equation}
  \dot{V} = x_3 N \frac{U V}{R + U} + x_6  N \frac{R V}{R + U}. \nonumber
\end{equation}
%------------------------
The derivative of $R(t)$ can be derived similarly. Let:
%------------------------
\begin{equation*}
  \begin{split}
  \int_0^{a_{1}} u(a,t_2)da = &\int_0^{a_{1}} u(a,t_1)da\\&+\int_{t_1}^{t_2} x_4e^{x_5(T_0-T)} dt\\&-\int_{t_1}^{t_2} u(a_1, t)dt\\&-\int_{t_1}^{t_2}  x_6 N \frac{R V}{R + U}.
  \end{split}
\end{equation*}
%------------------------
Because the much shorter life span of reticulocytes in comparison to mature red blood cells \cite{clark1990clinical} and the fact that the survival rate of reticulocytes to be close to 1 \cite{fonseca2016quantifying}, the hazard function, h(a), of reticulocyte is ignored. Substituting R, and rearrange the equations, we obtain: 
%------------------------
\begin{equation}
  R(t_2) - R(t_1) = \int_{t_1}^{t_2} x_4e^{x_5(T_0-T(t))} - u(a_1, t) - x_6 N \frac{R V}{R+U} dt. \nonumber
\end{equation}
%------------------------
Thus,
%------------------------
\begin{equation}
  \dot{R} =  x_4e^{x_5(T_0-T)} - u(a_1, t) - x_6 N \frac{R V}{R+U}. \nonumber
\end{equation}
%------------------------
Under the assumption that on average a small fixed percentage of reticulocyte are aging into mature red blood cell at any given time, 
%------------------------
\begin{equation}
  \dot{R} =  x_4e^{x_5(T_0-T)} - x_1R - x_6 N \frac{R I}{R+U}. \nonumber
\end{equation}
%------------------------
The derivative of $U(t)$ is also derived similarly. let: 
%------------------------
\begin{equation*}
  \begin{split}
  \int_{a_1}^{a_{max}} u(a,t_2)da = &\int_{a_{1}}^{a_{max}} u(a,t_1)da\\&+\int_{t_1}^{t_2} u(a_1,t) dt\\&-\int_{t_1}^{t_2} u(a_{max}, t)dt.\\&-\int_{t_1}^{t_2}\int_{a_1}^{a_{max}}h(a) u(a,t)dadt\\&-\int_{t_1}^{t_2}  x_3 N \frac{U V}{R+U}.
  \end{split}
\end{equation*}
%------------------------
Assuming that on average, a fixed percentage of red blood cells are removed due to random chance, substituting U then the equation becomes: 
%------------------------
\begin{equation}
  U(t_2) - U(t_1) = \int_{t_1}^{t_2}  u(a_1, t) - x_2U - x_6 N \frac{R I}{U+R} dt. \nonumber
\end{equation}
%------------------------
Thus: 
%------------------------
\begin{equation}
  \dot{U} =  x_1R - x_2U - x_6 N \frac{R I}{U+R}. \nonumber
\end{equation}
%------------------------
In conclusion, under the following explicit assumptions:
%------------------------
\begin{itemize}
	\item The change of infected red blood cell population and healthy red blood cell population is independent of the immune cell populations, 
    \item the erythropoiesis response does not have a delay, 
    \item on average, a fixed percentage of iRBCs are bursting at any given moment and a fixed percentage of reticulocytes are transitioning into mature red blood cells, 
\end{itemize}
%------------------------
the original PDE model \prettyref{eq:simplified system2} is simplified to the following system of ordinary differential equations that captures the change in the total infected red blood cell, red blood cell, and reticulocyte population: 
%------------------------
\begin{equation*}
	\begin{split}
      \dot{U} &= x_1 R  - x_2 U - x_3 N \frac{U I}{T},\\
      \dot{R} &= x_4 e^{x_5(T_0 - T)} - x_1 R - x_6 N \frac{R I}{T},\\
      \dot{I} &= x_3 N \frac{U I}{U + R} + x_6  N \frac{R I}{U + R}.
    \end{split}
\end{equation*}
%------------------------
Where U denote the concentration of healthy mature red blood cells (RBCs), R denotes healthy reticulocytes (RT) and I denote the concentration of infected red blood cells (iRBCs). Let $T = U + R$ and $T_0 = R_0 + U_0$ where $R_0$ and $U_0$ denote the steady state concentration of RTs and RBCs in the absence of malaria infection. N denote the average merozoites produced per infected red blood cell. $x_1 R$ describes the aging of RTs to become RBCs. $x_2 U$ denote the random removal of mature red blood cells. $x_3$ and $x_6$ denote the infection success rate of RBCs and RTs respectively. $x_4$ denotes the baseline production of RTs and the term $e^{x_5(T_0 - T)}$ describes the host erythropoiesis response. 

%%%%%%%%%%%%%%%%%%%%%%%%%%%%%%%%%%%%%%%%%%%%%%%%%%%%%%%
\section{Parameter Estimation}
%%%%%%%%%%%%%%%%%%%%%%%%%%%%%%%%%%%%%%%%%%%%%%%%%%%%%%%
Under the assumption of steady state, where $R_0$ is the steady state concentration of RTs and $U_0$ is the steady state concentration of RBCs, the following equality is established: 
%------------------------
\begin{equation}
  x_4 = x_1 R_0 = x_2 U_0. \nonumber
\end{equation}	
%------------------------
The average number of merozoite produced is set to be 20. Fonseca \textit{et al.} \cite{fonseca2016quantifying} estimated that the baseline production of RTs in \textit{M mulatta}, $x_4$, is $192,00$ cells per day \cite{fonseca2016quantifying}. $R_0$ and $U_0$ are estimated for each subjects using the average of the red blood cell and reticulocytes counts during the first ten days of the experiment where no iRBCs were detected. The lower bound and upper bound of the other three parameters, $x_3$, $x_6$ and $x_5$ were also estimated. All three parameters are positive. The upper bound of $x_5$ was set to allow a maximum of 2-fold increase in RTs production in each subject. Upper bound of $x_3$ and $x_6$ was set to be $0.05$ and $1$ respectively. When $x_3 > 0.05$ and/or $x_6 > 1$, the system becomes numerically unstable to solve due to stiffness using odesolver45 in MATLAB environment. The upper bound of $x_3$ and $x_6$ also ensures that the $\dot{I} < 0.5NI$ throughout the simulation, meaning that the number of infected red blood cells produced over a two day period can not exceed the number of merozoites produced. 

A multiple objective genetic algorithm \cite{deb2005multi} was used to estimate the parameters ($x_3,x_6,x_5$) for each subjects using the estimated lower and upper bound of each parameter. The model was fitted to the cellular data with starting date corresponding to the first appearance of iRBCs and ending date corresponding to time point 2. The two objectives minimized were the average percent error (APE) of predicted RTs and iRBCs concentration. APE have the following form for a specific variable such as RTs: 
%------------------------
\begin{equation}
  APE_{retic} = \frac{100}{n} \sum_{i=1}^n \frac{r_i - R_i}{R_i}, \nonumber
\end{equation}
%------------------------
where $r_i$ is the observed RTs concentration at time point $i$ and $R_i$ is the model predicted RTs concentration. Contrary to single objective minimization, a Pareto front for both objectives is estimated during each application of the genetic algorithm. The Pareto front refers to a set of possible values of both object function such that the decrease in one objective function necessitates the increase of the other \cite{deb2005multi}. Each application of a genetic algorithm is terminated when the change in the estimated Pareto front is less than the predefined tolerance ($0.0001$). Due to the stochastic nature of the genetic algorithm, 1,000 runs were applied to estimate the parameters for each subject. $> 99\%$ of the application of genetic algorithm terminated due to the convergence of Pareto front, the rest did not converge during the maximum allowed run time of $60$ seconds. 

%%%%%%%%%%%%%%%%%%%%%%%%%%%%%%%%%%%%%%%%%%%%%%%%%%%%%%%
\section{Results}
%%%%%%%%%%%%%%%%%%%%%%%%%%%%%%%%%%%%%%%%%%%%%%%%%%%%%%%
\subsection{Result Overview}
%%%%%%%%%%%%%%%%%%%%%%%%%%%%%%%%%%%%%%%%%%%%%%%%%%%%%%%
%------------------------
\begin{figure}[!ht]
	\centering
	\includegraphics[width = 0.5\textwidth]{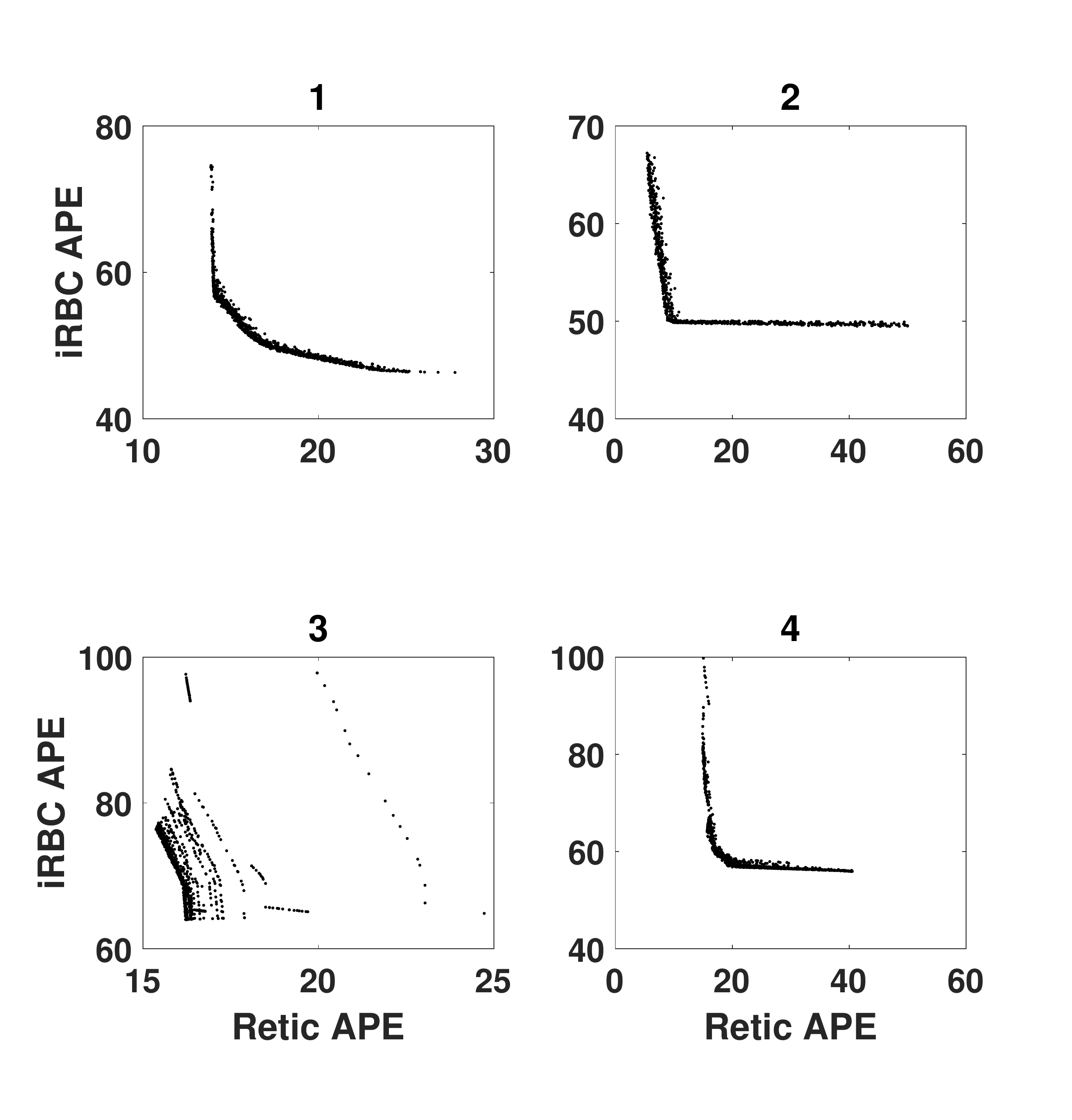}
	\caption[Pareto Front of Parameter Estimation]
    {Estimated Pareto Front of Average Percentage Error (APE) of iRBCs and RTs for each of the four subjects.}
	\label{fig:Pareto}
\end{figure}
\begin{figure}[!ht]
	\centering
	\includegraphics[width = 0.5\textwidth]{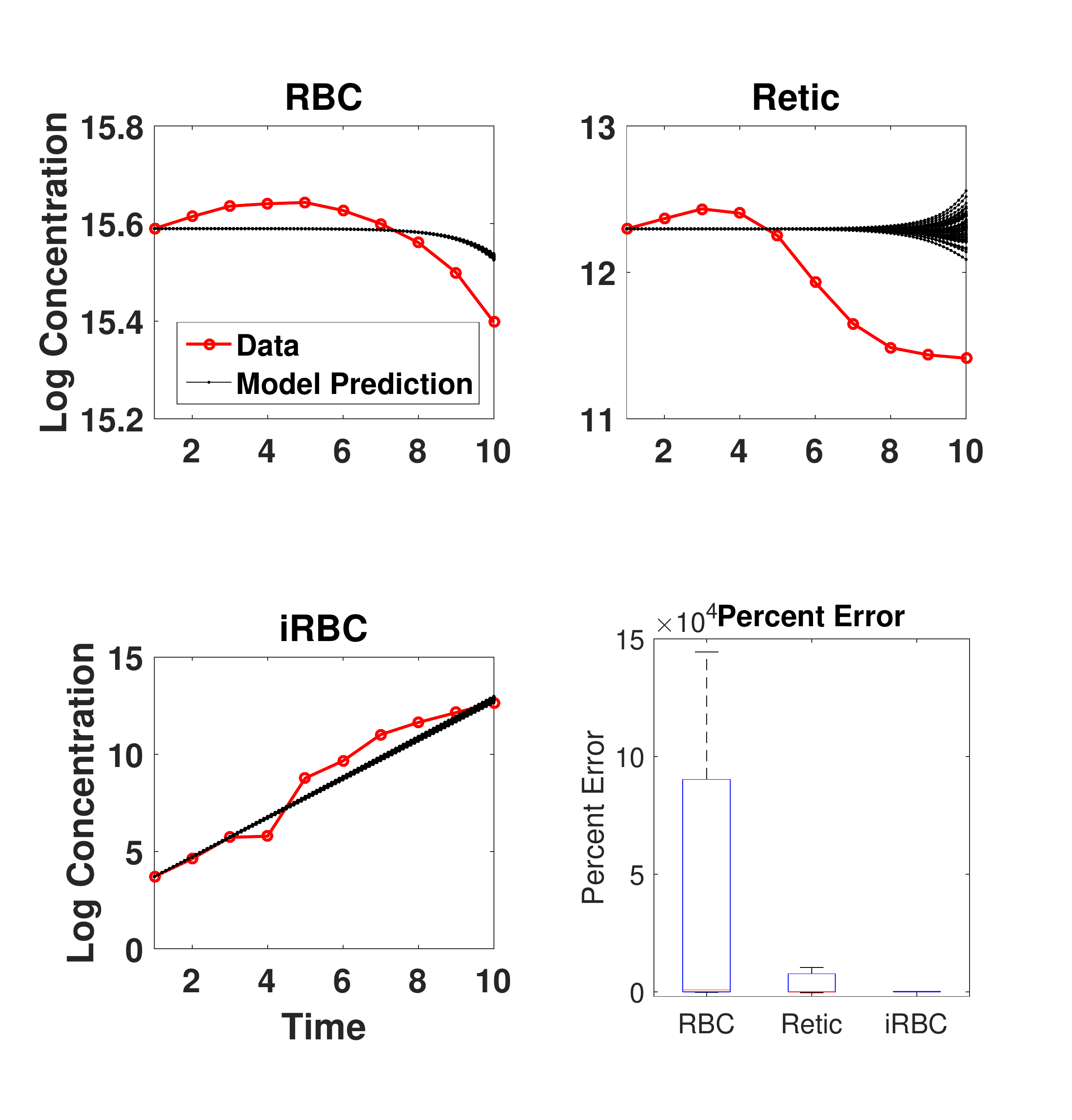}
	\caption[Subject 1 Simulation Result]
    {Simulation of the top 100 models with the lowest iRBCs APE and distribution of percent error of subject 1}
	\label{fig:MD-Fit S1}
\end{figure}
\begin{figure}[!ht]
	\centering
	\includegraphics[width = 0.5\textwidth]{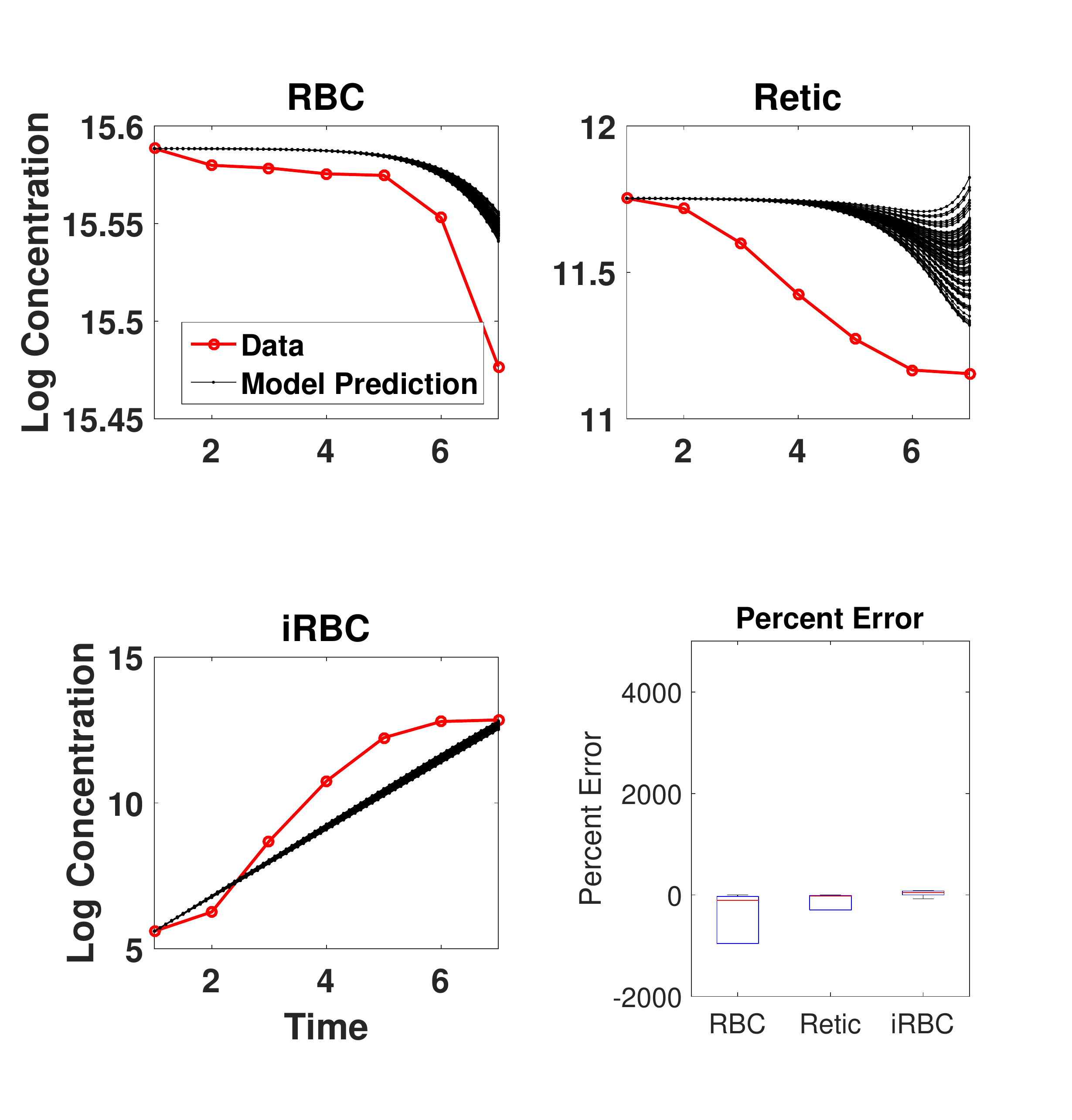}
	\caption[Subject 2 Simulation Result]
    {Simulation of the top 100 models with the lowest iRBCs APE and distribution of percent error of subject 2}
	\label{fig:MD-Fit S2}
\end{figure}
\begin{figure}[!ht]
	\centering
	\includegraphics[width = 0.5\textwidth]{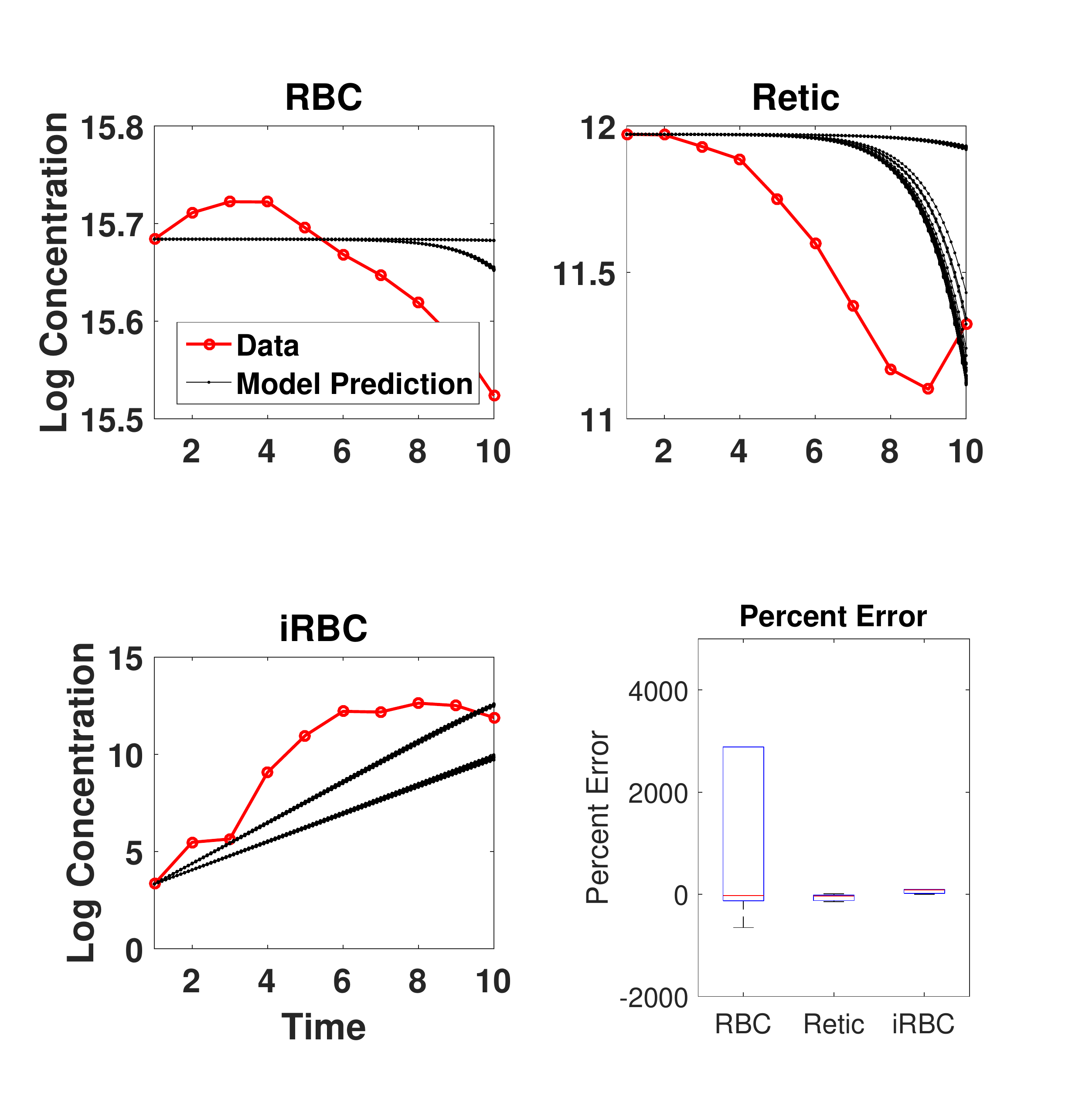}
	\caption[Subject 3 Simulation Result]
    {Simulation of the top 100 models with the lowest iRBCs APE and distribution of percent error of subject 3}
	\label{fig:MD-Fit M1}
\end{figure}
\begin{figure}[!ht]
	\centering
	\includegraphics[width = 0.5\textwidth]{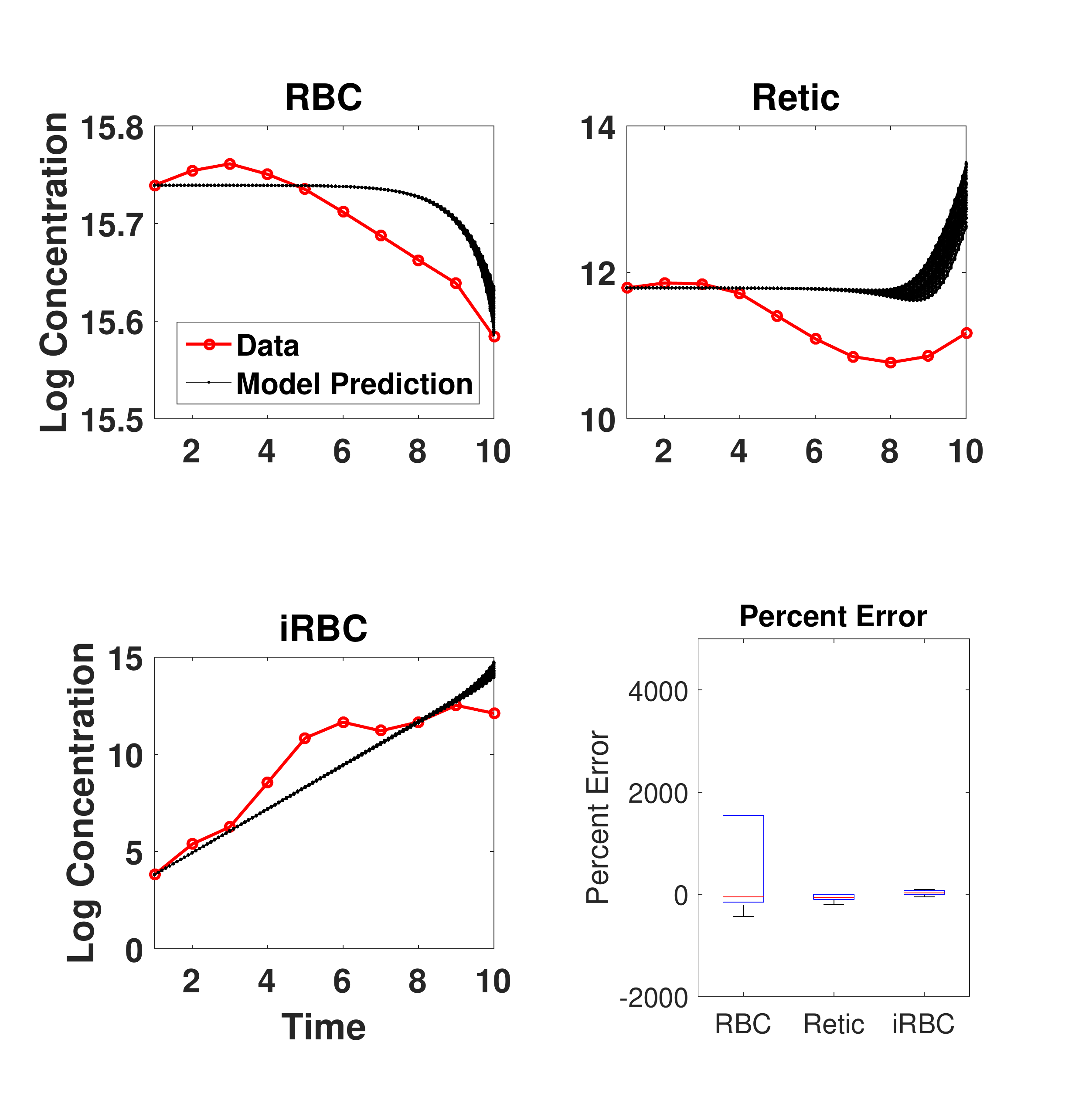}
	\caption[Subject 4 Simulation Result]
    {Simulation of the top 100 models with the lowest iRBCs APE and distribution of percent error of subject 4}
	\label{fig:MD-Fit M2}
\end{figure}

%------------------------
The Pareto front of iRBC APE and RT APE for each subject are shown in Figure \ref{fig:Pareto}. For subject 1, 2 and 4, there exist a sub-region on the Pareto front such that the iRBC APE and RT APE have a negative linear relationship, which means that $iRT APE + RT APE$ is close to constant. The Pareto front estimated for subject 3 does not contain such a region, indicating a lack of model fit for that specific subject. The iRBC APE for all four subjects has a range of $(45\% \sim 90\%)$ where the RT APE have a range of ($15\% \sim 50\%$). The simulation of the top 100 model with the lowest iRBCs APE are shown in Figure (\ref{fig:MD-Fit S1}, \ref{fig:MD-Fit S2}, \ref{fig:MD-Fit M1}, \ref{fig:MD-Fit M2}). 
%------------------------
%------------------------
%------------------------
%------------------------
%------------------------

%%%%%%%%%%%%%%%%%%%%%%%%%%%%%%%%%%%%%%%%%%%%%%%%%%%%%%%
\subsection{Quantification of The Preferential Infection of Reticulocyte by \textit{P. cynomolgi}}
%%%%%%%%%%%%%%%%%%%%%%%%%%%%%%%%%%%%%%%%%%%%%%%%%%%%%%%
\begin{figure}[!ht]
	\centering
	\includegraphics[width = 0.5\textwidth]{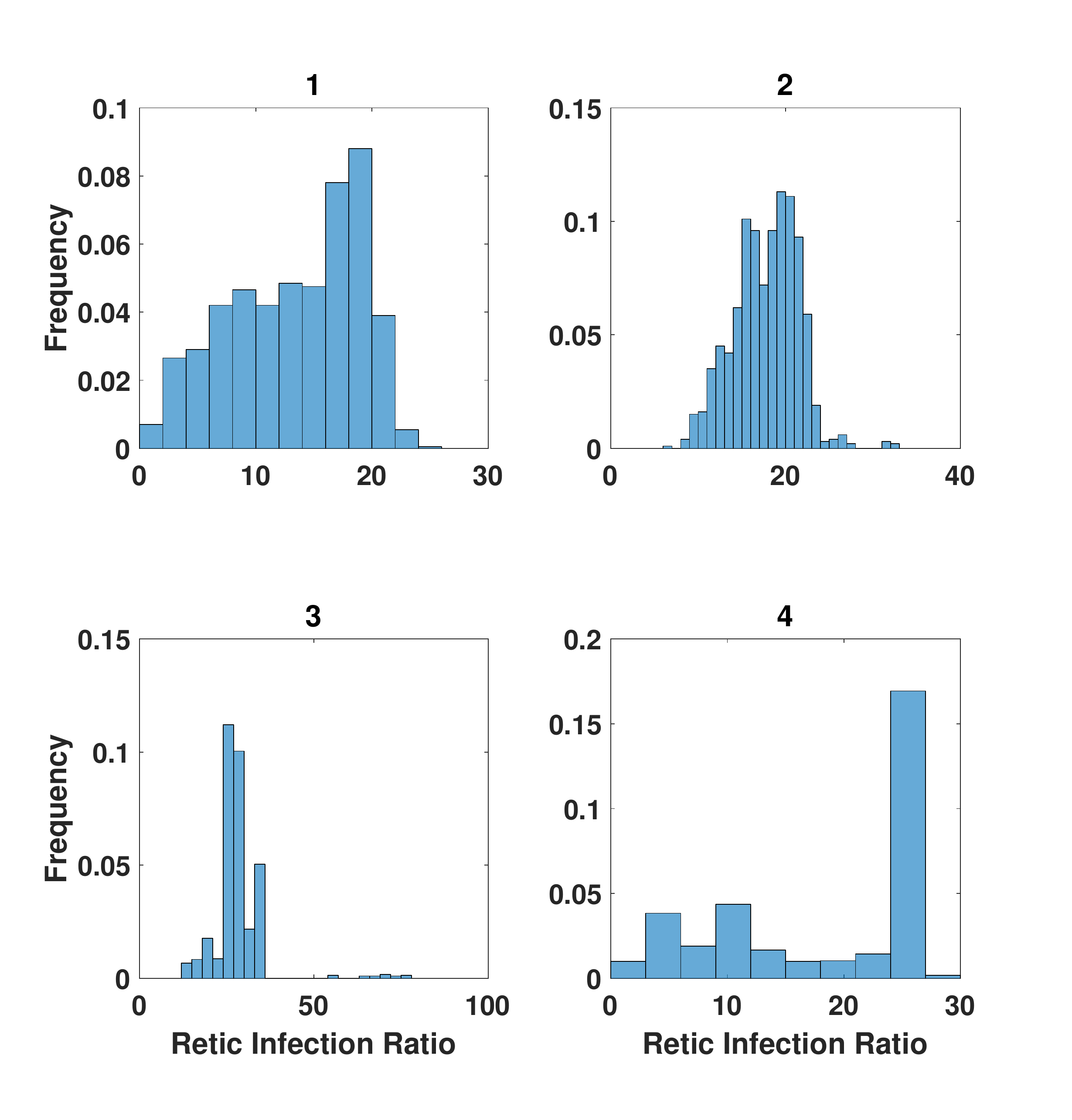}
	\caption[Distribution of $\frac{x_6}{x_3}$]
    {Distribution of $\frac{x_6}{x_3}$ for subjects 1 ~ 4.}
	\label{fig:ReticInfection}
\end{figure}
\newpage 

To quantify the preferential infection of RTs by \textit{P cynomolgi}, the ratio of $x_6$ to $x_3$ in the top 100 model with the lowest iRBCs APE for each subject was calculated. The ratio $\frac{x_6}{x_3}$ is interpreted as the likelihood of \textit{P cynomolgi} infecting RTs over infecting RBCs. The distribution of $\frac{x_6}{x_3}$ is shown in Figure \ref{fig:ReticInfection}. The means of the four distribution are 13, 17, 28 and 18 respectively.

%%%%%%%%%%%%%%%%%%%%%%%%%%%%%%%%%%%%%%%%%%%%%%%%%%%%%%%
\subsection{Quantification of Removal of Healthy Red Blood Cell (hRBC) }
%%%%%%%%%%%%%%%%%%%%%%%%%%%%%%%%%%%%%%%%%%%%%%%%%%%%%%%
%------------------------
\begin{figure}[!ht]
	\centering
	\includegraphics[width = 0.5\textwidth]{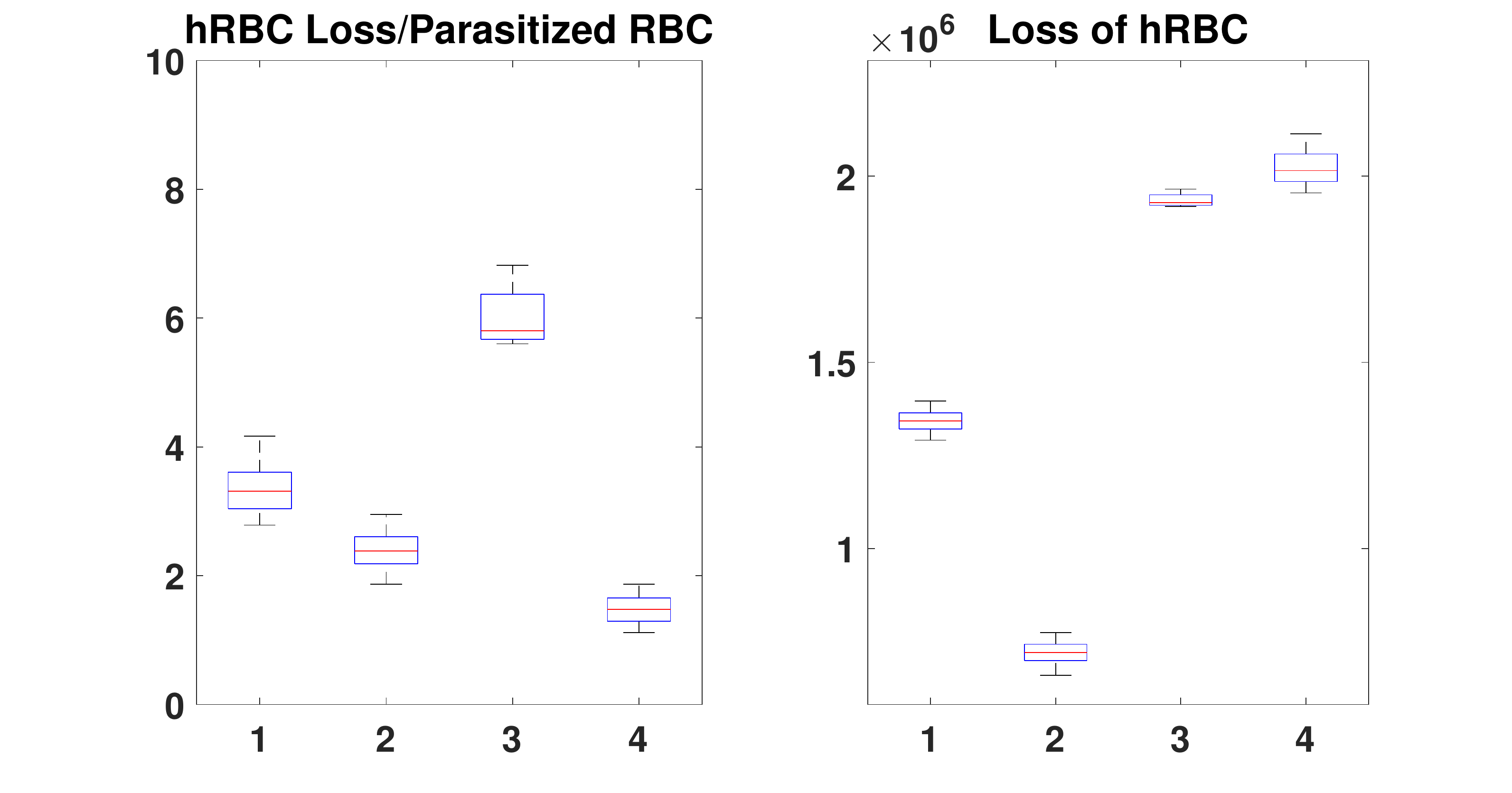}
	\caption[Estimated Total Loss of hRBC]
    {Estimated total loss of hRBC and ratio of loss of hRBC and parasitized RBC in subject 1-4}
	\label{fig:RBCLoss}
\end{figure}
%------------------------
\begin{figure}[!ht]
	\centering
	\includegraphics[width = 0.5\textwidth]{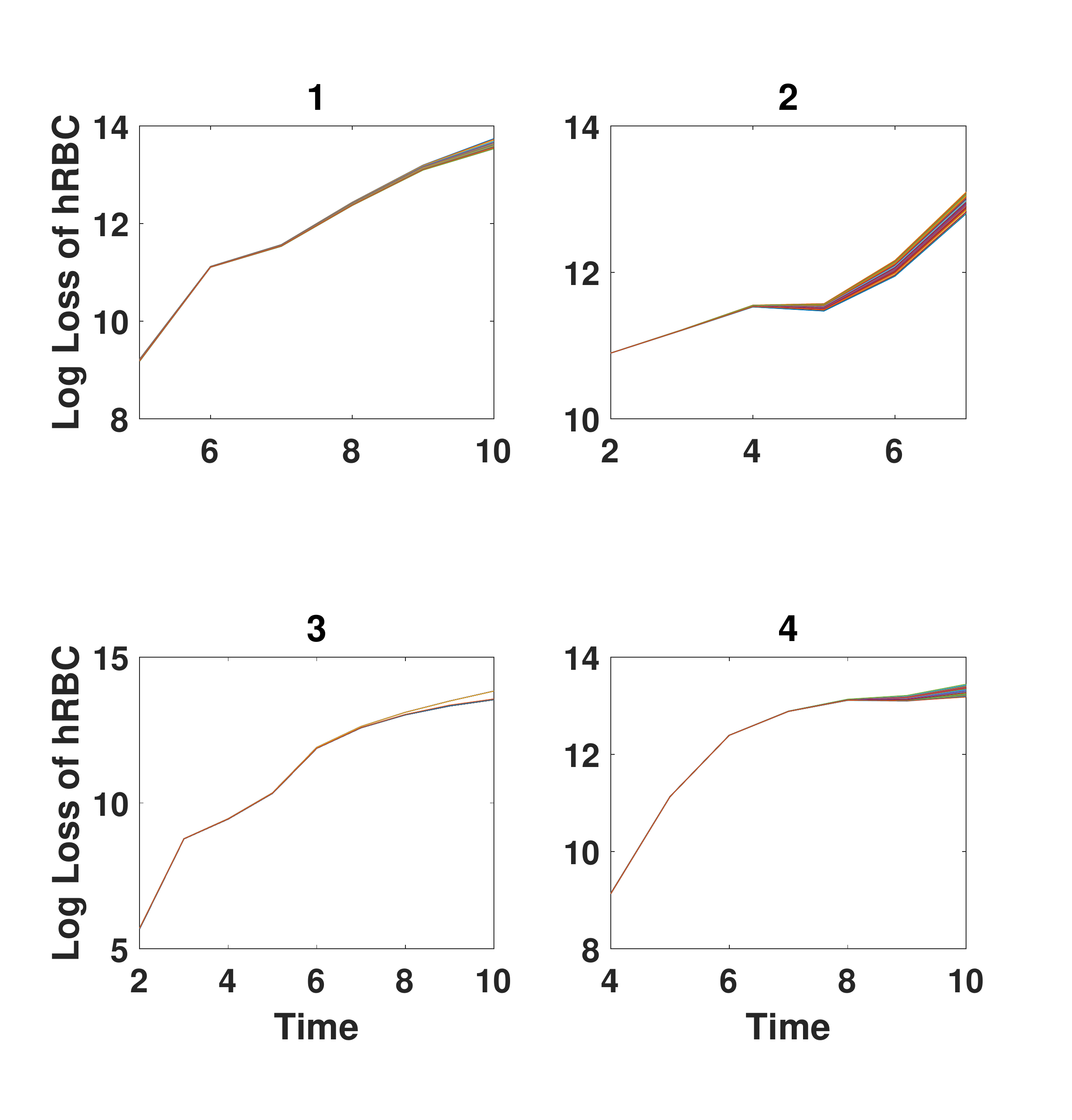}
	\caption[Time Series of Estimated Loss of hRBCs]
    {Time series of estimated loss of hRBCs.}
	\label{fig:RBCLossTimeSeries}
\end{figure}
%------------------------
\begin{figure}[!ht]
	\centering
	\includegraphics[width = 0.5\textwidth]{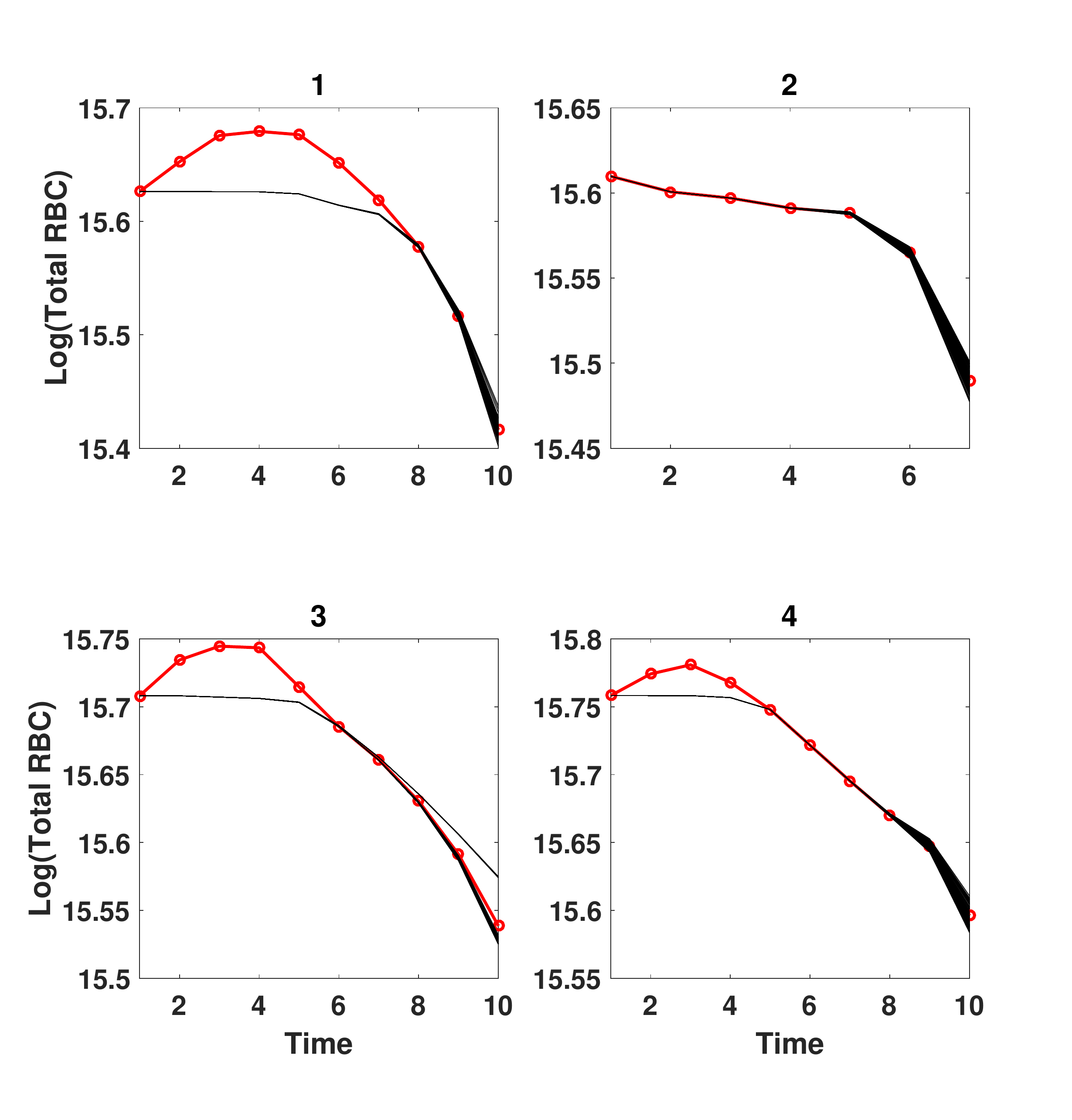}
	\caption[Time Series of hRBC Loss Adjusted Model]
    {Time series of model predicted total red blood cell adjusted by the mean of estimated loss of hRBC.}
	\label{fig:Adjusted Total RBC}
\end{figure}
%------------------------

The loss of healthy RBC (hRBC) is also calculated based on the 100 model prediction with the lowest iRBC APE. Total Loss of hRBC and Ratio of loss of hRBC and parasitized RBC are shown in Figure \ref{fig:RBCLoss}. On average, 1.5 million hRBCs per $\mu$l are cleared by the host through out the onset of the disease till acute primary infection. Furthermore, our model predicted that for each RBC parasitized, the host removes 3 hRBC. The quantification of the removal of hRBCs over the entire infection is shown in Figure \ref{fig:RBCLossTimeSeries}. The model fit of RBC and retic Population after adjusting for removal of hRBCs are shown in Figure \ref{fig:Adjusted Total RBC}. 

%%%%%%%%%%%%%%%%%%%%%%%%%%%%%%%%%%%%%%%%%%%%%%%%%%%%%%%
\section{Correlation and Enrichment Analysis of Rate of hRBC Removal}
%%%%%%%%%%%%%%%%%%%%%%%%%%%%%%%%%%%%%%%%%%%%%%%%%%%%%%%

\begin{figure}[!ht]
	\centering
	\includegraphics[width = 0.5\textwidth]{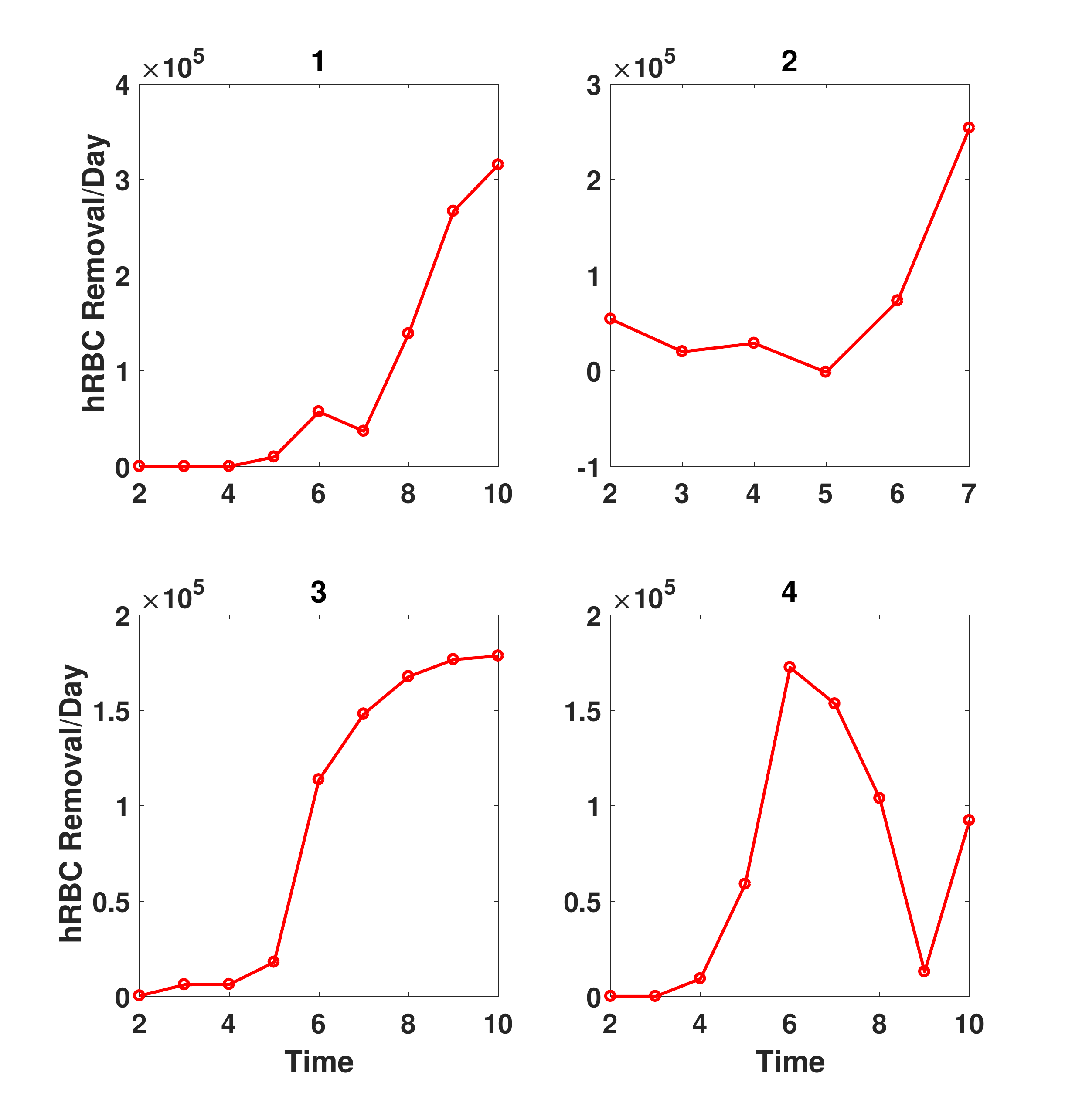}
	\caption[hRBC Removal Rate]
    {Time series of model predicted hRBC removal rate for all four subjects.}
	\label{fig:hRBC Removal Rate}
\end{figure}

\begin{figure}[!ht]
	\centering
	\includegraphics[width = \textwidth]{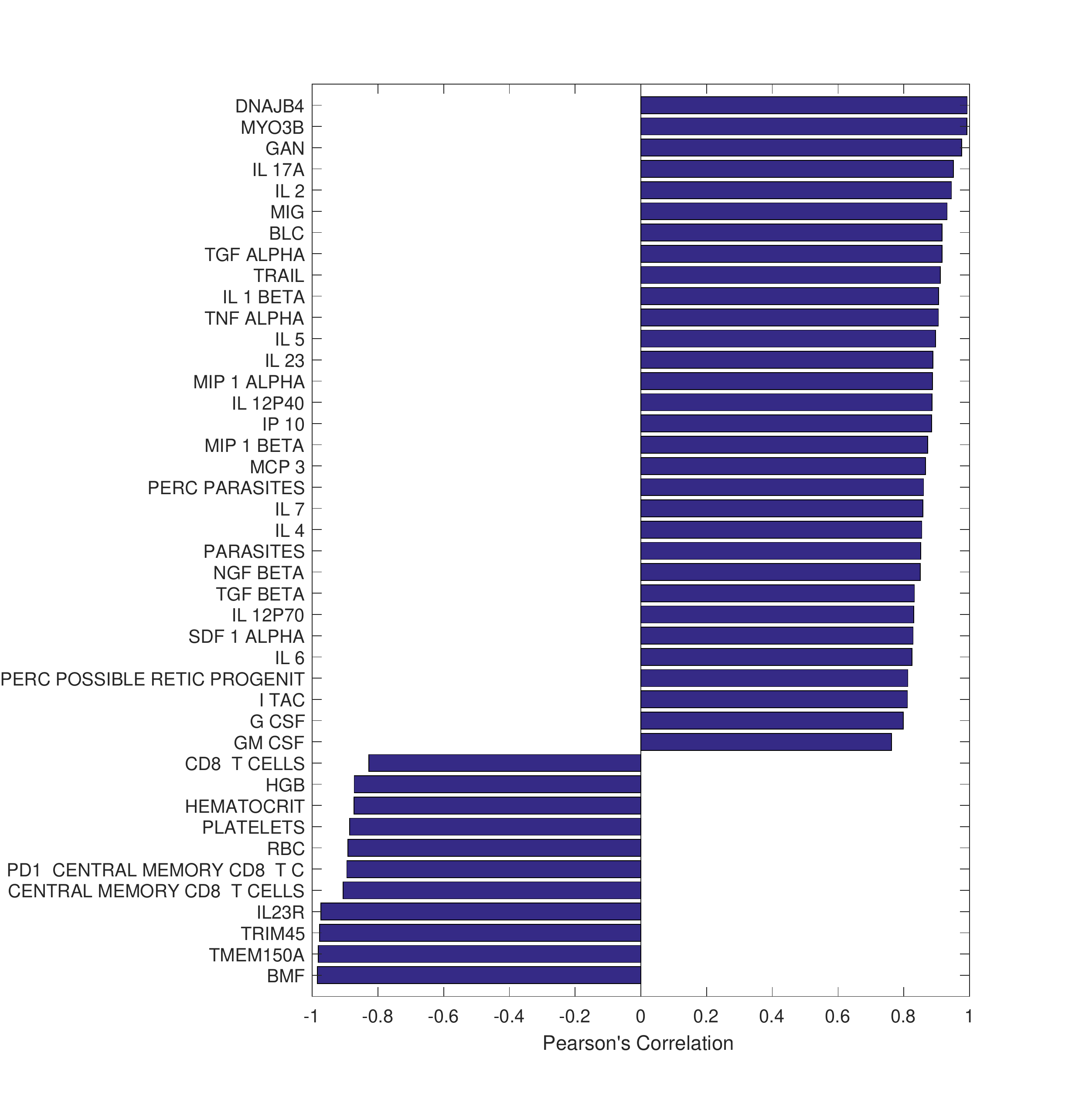}
	\caption[Correlation of hRBC Removal Rate]
    {Transcripts, immune cell population and cytokines that are highly correlated (q-val $\leq 0.05$) with hRBC removal rate.}
	\label{fig:hRBC Corr}
\end{figure}

The rate of hRBC removal was estimated for all four subjects (Fig  \ref{fig:hRBC Removal Rate}) based on the estimated loss of hRBCs (Fig \ref{fig:RBCLossTimeSeries}). Pearson's correlation between hRBC removal rate and their corresponding transcript, immune cell and cytokine abundance was calculated. The p-value of the correlation was adjusted for false discovery rate and the transcript, immune cell and cytokine exhibiting significant correlation with hRBC removal rate (q-val $\leq 0.05$) are shown in Fig \ref{fig:hRBC Corr}. Several pro-inflammatory ctokines such as IL6 and IL1B are positively correlated with the rate of hRBC removal. On the other hand, CD8 T cell population displayed significant negative correlation with hRBC removal rate.  

Only seven genes: MYO3B, GAN, DNAJB4, TRIM45, TMEM150A, IL23R and BMF have shown significant correlation with the rate of hRBC removal. To fully explore the association of transcriptome change and rate of hRBC removal, Gene Set Enrichment Analysis \cite{subramanian2005gene} was applied to the correlation ranked gene lists. The most significantly enriched GO gene sets and pathways are shown in table \ref{table:GO Gene Set Enrichment} and table \ref{table:Pathway Enrichment}. Innate immune related gene sets and pathways are positively correlated with hRBC removal rate, which corresponds to the positive correlation between hRBC removal and inflammatory cytokines. On the other hand, transcripts that have negative correlation with hRBC removal are enriched in RNA and protein processing related pathways. 
%------------------------
%------------------------
%------------------------

\begin{table}
	\footnotesize % \tiny
	\centering
	\caption
	{Enrichment of GO Gene Sets} \label{table:GO Gene Set Enrichment} 
	\begin{tabular}{ll}
		\addlinespace
		\toprule
		
		Positive Enriched GO Gene Sets & Negative Enriched GO Gene Sets \\
		\midrule
  INNATE IMMUNE RESPONSE&	  NCRNA PROCESSING\\
  IMMUNE EFFECTOR PROCESS&	  NCRNA METABOLIC PROCESS\\
  ACTIN FILAMENT BASED PROCESS&	  RIBOSOME BIOGENESIS\\
  REGULATION OF BODY FLUID LEVELS&	  RRNA METABOLIC PROCESS\\
  WOUND HEALING&	  RIBONUCLEOPROTEIN COMPLEX BIOGENESIS\\
  RESPONSE TO BACTERIUM&	  TRANSLATIONAL INITIATION\\
  DEFENSE RESPONSE TO OTHER ORGANISM&	  AMIDE BIOSYNTHETIC PROCESS\\
  INFLAMMATORY RESPONSE&	  PEPTIDE METABOLIC PROCESS\\
  ACTIVATION OF IMMUNE RESPONSE&	  MULTI ORGANISM METABOLIC PROCESS\\
  REGULATION OF RESPONSE TO WOUNDING&	  NONSENSE MEDIATED DECAY\\
  HEMOSTASIS&	  PROTEIN LOCALIZATION\\
  REGULATED EXOCYTOSIS&	  PROTEIN LOCALIZATION TO ENDOPLASMIC RETICULUM\\
  SECRETION&	  PROTEIN TARGETING TO MEMBRANE\\
  RESPONSE TO WOUNDING&	  TRNA METABOLIC PROCESS\\
  CELLULAR RESPONSE TO NITROGEN COMPOUND&	  RNA CATABOLIC PROCESS\\
  ENDOCYTOSIS&	  ORGANIC CYCLIC COMPOUND CATABOLIC PROCESS\\
  RESPONSE TO VIRUS&	  TRNA PROCESSING\\
  CELL CELL ADHESION&	  MITOCHONDRIAL TRANSLATION\\
  REGULATION OF INFLAMMATORY RESPONSE&	  TRANSLATIONAL ELONGATION\\

		\bottomrule
	\end{tabular}
\end{table} 
\newpage

\begin{table}
	\footnotesize % \tiny
	\centering
	\caption
	{Enrichment of Pathways} \label{table:Pathway Enrichment} 
	\begin{tabular}{ll}
	\addlinespace
		\toprule
		
		Positive Enriched Pathways & Negative Enriched Pathways\\
		\midrule
SYSTEMIC LUPUS ERYTHEMATOSUS&3 UTR MEDIATED TRANSLATIONAL REGULATION\\
  HEMOSTASIS&TRANSLATION\\
  CYTOKINE SIGNALING IN IMMUNE SYSTEM&PEPTIDE CHAIN ELONGATION\\
 CHEMOKINE SIGNALING PATHWAY&RIBOSOME\\
  INTERFERON ALPHA BETA SIGNALING&INFLUENZA VIRAL REPLICATION\\
PID PDGFRB PATHWAY& PROTEIN TARGETING TO MEMBRANE\\
 REGULATION OF ACTIN CYTOSKELETON&	  NONSENSE MEDIATED DECAY \\
  PLATELET ACTIVATION SIGNALING AND AGGREGATION&	  INFLUENZA LIFE CYCLE\\
  AMYLOIDS&	  METABOLISM OF PROTEINS\\
PID VEGFR1 2 PATHWAY&	  METABOLISM OF RNA\\
  RNA POL I PROMOTER OPENING&	  FORMATION OF THE TERNARY COMPLEX\\
NABA MATRISOME ASSOCIATED&	  ACTIVATION OF THE MRNA BINDING TO 43S\\
  RESPONSE TO ELEVATED PLATELET CYTOSOLIC CA2&	  METABOLISM OF MRNA\\
NABA SECRETED FACTORS&	  MITOCHONDRIAL PROTEIN IMPORT\\
  INTERFERON SIGNALING&	 RNA POLYMERASE\\
  RHO PATHWAY&	  RNA POL III TRANSCRIPTION INITIATION \\
 LEUKOCYTE TRANSENDOTHELIAL MIGRATION&RESPIRATORY ELECTRON TRANSPORT ATP SYNTHESIS\\
 FOCAL ADHESION&	 PYRIMIDINE METABOLISM\\
 JAK STAT SIGNALING PATHWAY&TCA CYCLE AND RESPIRATORY ELECTRON TRANSPORT\\
		\bottomrule
	\end{tabular}
\end{table} 
\newpage

%%%%%%%%%%%%%%%%%%%%%%%%%%%%%%%%%%%%%%%%%%%%%%%%%%%%%%%
\section{Empirical Model Adjustment}
\begin{figure}[!ht]
	\centering
	\includegraphics[width = 0.5\textwidth]{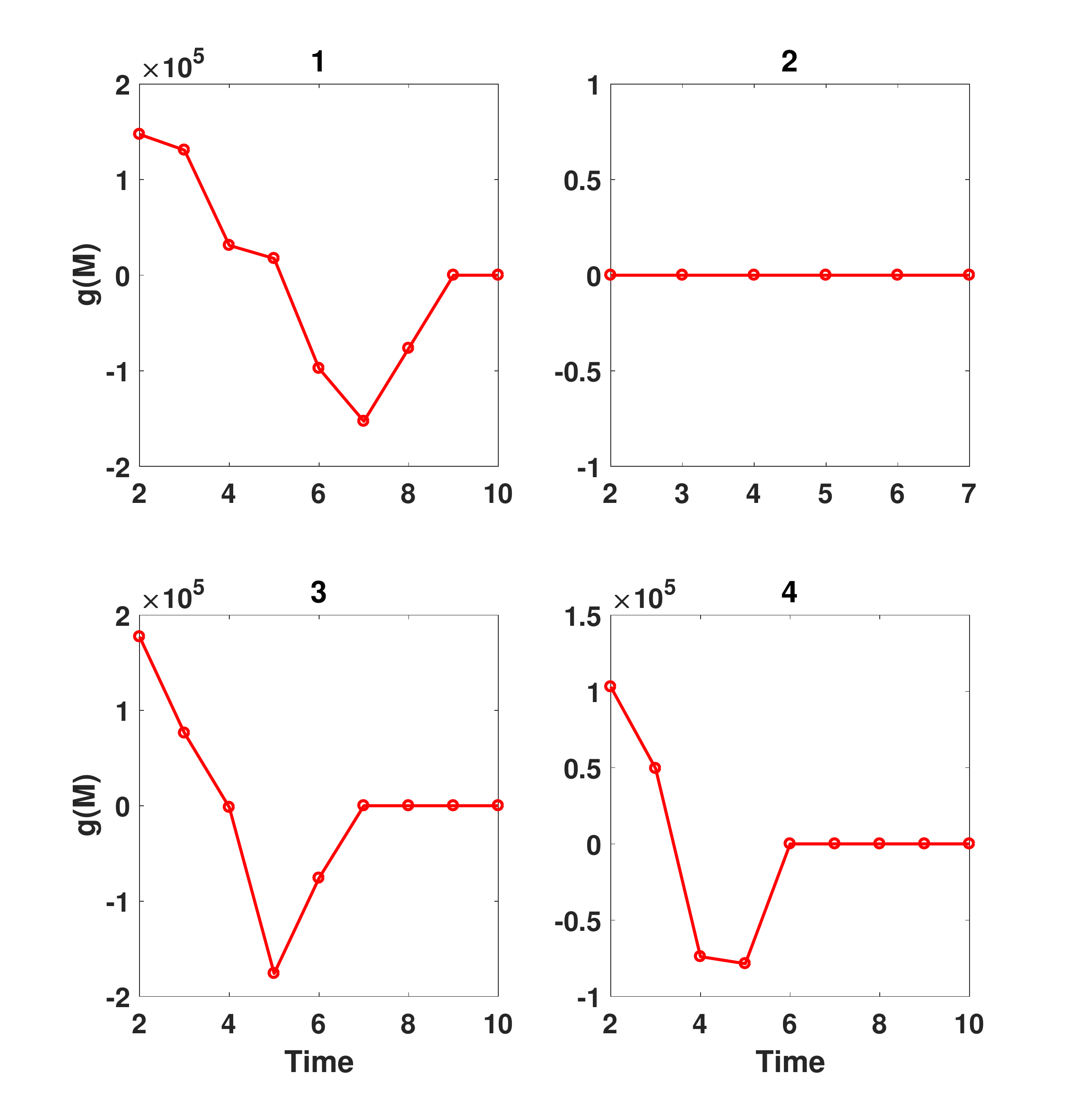}
	\caption[Time Series of g(M)]
    {Time series of g(M) for subject 1 through 4.}
	\label{fig:g(M)}
\end{figure}

\begin{figure}[!ht]
	\centering
	\includegraphics[width = 0.7\textwidth]{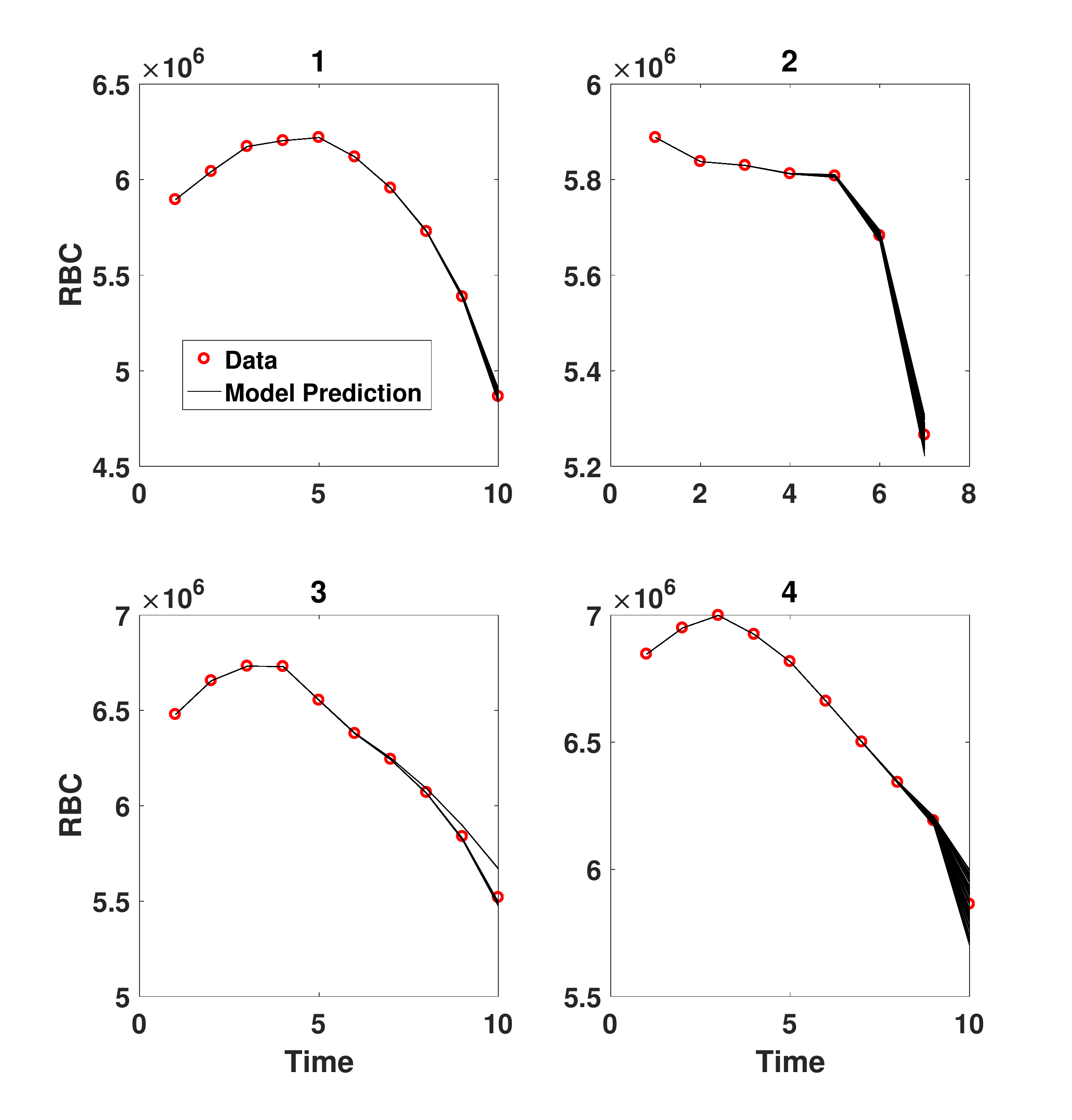}
	\caption[Final Adjusted Model]{Final adjusted model prediction of RBC population for subjects 1 through 4.}
	\label{fig:Final Model}
\end{figure}

%%%%%%%%%%%%%%%%%%%%%%%%%%%%%%%%%%%%%%%%%%%%%%%%%%%%%%%
The simulation of the top 100 model with the lowest iRBC APE for subject 1,2 and 3 (Figure \ref{fig:MD-Fit S1}, \ref{fig:MD-Fit S2}, \ref{fig:MD-Fit M1}, \ref{fig:MD-Fit M2}) failed to capture the increase of RBC population during the early stages of malaria infection. Considering that the difference between RBC population and steady state RBC population is minimal during the early stage of malaria infection, this discrepancy suggest that RBCs are released during the early stage of blood stage malaria. Additionally, our correlation analysis identified several cell types, cytokines and transcripts (q-value $\leq 0.05$) that are linearly correlated with hRBC removal rate. This finding suggest the possibility of using these entities to predict hRBC removal rate. Taking these findings into consideration, our original model can be further expanded to: 
%------------------------
\begin{equation*}
	\begin{split}
      \dot{U} &= x_1 R  - x_2 U - x_3 N \frac{U I}{T} - \sum \gamma_i E_i + g(M),\\
      \dot{R} &= x_4 e^{x_5(T_0 - T)} - x_1 R - x_6 N \frac{R I}{T},\\
      \dot{I} &= x_3 N \frac{U I}{U + R} + x_6  N \frac{R I}{U + R},
    \end{split}
\end{equation*}
%------------------------
 is linearly explained by a combination of the abundances of inflammatory cytokines. Where each $E_i$ denote the abundance of a cytokine exhibiting linear correlation with the rate of hRBC removal (Fig:  \ref{fig:hRBC Corr}) and each $\gamma_i$ quantifies the linear dependency. The term $g(M)$ denote the increase of RBC population during the early stage of malaria infection, that is dependent on some molecular quantity $M$, possibly pathogen related. The estimated form of $g(M)$ are shown in Figure \ref{fig:g(M)} and the post-hoc fitted model is shown in Figure \ref{fig:Final Model}. 
%------------------------
%------------------------
%------------------------

%%%%%%%%%%%%%%%%%%%%%%%%%%%%%%%%%%%%%%%%%%%%%%%%%%%%%%%
\section{Discussion}
%%%%%%%%%%%%%%%%%%%%%%%%%%%%%%%%%%%%%%%%%%%%%%%%%%%%%%%
In this chapter, a simplified Ordinary Differential Equation model was fitted to the time series data collected from the MaHPIC experiment. Our simplified model was derived from the earlier PDE system under three explicit assumptions. The simplification of the original model reduced the unknown parameters to 3 parameters with well-defined lower and upper bounds. 

The model was fitted using multiple objective genetic algorithm and the Pareto front estimated (Fig: \ref{fig:Pareto}) demonstrate that the simplified model describes the dynamics of RTs much better than that of iRBCs as shown by the 2-fold difference in the range of RT APE and iRBC APE. The iRBC APE for all four subjects has a range of $(45\% \sim 90\%)$ where as the RT APE have a range of ($15\% \sim 20\%$). This result suggests that the immune function omitted in our model have a large impact on parasite population during the primary infection. 

The best-fitted models were used to quantify the preferential infection of RTs by \textit{P. cynomolgi}. Our model predicts that, on average, \textit{P. cynomolgi} merozoites are 20 times more likely to infect RTs than RBC (Fig \ref{fig:ReticInfection}). The experimental verification of our prediction is difficult due to the lack of an \text{in vitro} system to study \textit{P. cynomolgi}, but our prediction provides a general range of preferential infection of RTs, which can be utilized in future experimental designs and modeling studies. 

Additionally, in all the best fit models (Fig \ref{fig:MD-Fit S1}, \ref{fig:MD-Fit S2}, \ref{fig:MD-Fit M1}, \ref{fig:MD-Fit M2}), our prediction of RBC concentration and RT concentration are higher than observed. This over estimation of RBC and RT concentration constitutes the majority of APE. Furthermore, our iRBC concentration predictions are on average lower than the observed value. We utilized this fact to estimate the lower bound of the amount of hRBCs removed by the host. Our model estimates that at least $50\% \sim 80\%$ of RBC loss during the primary infection of \textit{P. cynomolgi} are due to the removal of hRBCs \ref{fig:RBCLoss}. Finally, our model shows that the speed at which hRBCs are removed increases throughout the infection. 

Using the estimated hRBC removal rate, a correlation study was conducted to identify transcript, immune cell and cytokine abundance that has significant correlation with hRBC removal rate. Interestingly,  innate immune related gene sets (Interferon Response, Rho Pathway and JAK-STAT pathway \textit{et al.}) along with pro-inflammatory cytokines (IL-1B, IL-6 \textit{et al.}) displayed significant positive correlation with hRBC removal. The association of severe malaria anemia with pro-inflammatory response has long been studied \cite{perkins2011severe}; our analysis provides a list of possible cytokine biomarkers for the estimation of host clearance of hRBCs. Due to the experimental constraint where cytokine profiling was only conducted during peak parasitemia, our analysis lacks the resolution to provide a mechanistic explanation for the correlation between pro-inflammatory cytokines and host removal of hRBC. Furthermore, the observation that inflammation related genes and cytokines are differentially up-regulated in the severe subjects (Subject 1 and 2), along with the fact that severe subjects have a higher rate of hRBC removal during peak parastemia, suggest the possible role of inflammation associated hRBC clearance and clinical severity. 

In conclusion, we have demonstrated that a simplified model with only three unknown parameters can be used to predict RT concentration with an APE of ($15\% \sim 20\%$). Despite the model's relatively poor performance at predicting iRBC dynamics (APE $50\% \sim 70\%$), it can be used to estimate the preferential infection of RTs and hRBC removal during malaria infection. The estimation of the hRBC removal rate using our model along with the down-stream enrichment analysis reveals associations of hRBC removal and both the inflammatory response and CD 8 T-cell response. Application of this model to more time series data sets of malaria infection involving a variety of malaria species is necessary to validate our findings. 

%%%%%%%%%%%%%%%%%%%%%%%%%%%%%%%%%%%%%%%%%%%%%%%%%%%%%%%
\section*{\textbf{APPENDIX}}
%%%%%%%%%%%%%%%%%%%%%%%%%%%%%%%%%%%%%%%%%%%%%%%%%%%%%%%
The parameters, functions, and variables described by Yan \textit{et al.} \cite{yan2015mathematical} are presented here for completeness. The reader is encouraged to read that article for a detailed description of each element. 
\begin{center}
		{\scriptsize
		\centering
    \begin{tabular}{ | l | p{5cm} | l |  }
			\hline
			          			& Description & Unit 		 \\ \hline
			\multicolumn{3}{ |c| }{Independent Variables} \\
			\hline
			
			$t$							&Time.         & $Day$		 \\ \hline
			$a$							&Age of RBC. & $Day$  \\ \hline
			$\alpha$				&Age of iRBC. & $10^-1$$Day$  \\ \hline
			
			\multicolumn{3}{ |c| }{Dependent Variables} \\
			\hline
			
			$u(a, t)$				&Circulating RBC age distribution. &$cells/ul/day$ \\ \hline
			$v(\alpha, t)$   &iRBC age distribution. &$cells/ul/day$ \\ \hline
			$\varphi(t)$		&Circulating RBC concentration. &$cells/ul$  \\ \hline
			$w_{i}(t)$			&Circulating innate immune cell concentration. &$cells/ul$  \\ \hline
			$s_{i}(t)$			&Circulating adaptive immune cell concentration. &$cells/ul$  \\ \hline
			\multicolumn{3}{ |c| }{Functions} \\ \hline

			$h(a)$          & Percentage of RBC that leaves circulation per day, usually a constant. 			  & 1/day	 	   \\ \hline
			$f(t, \varphi_{t})$ &Rate at which RBC enters circulation. 		& $cells/ul/day$ \\ \hline
			$r(a)$ 					& Success rate of merozoite invading RBC of age. $a$		& $1/day$	   \\ \hline

			$p(u(a, t))$		& Probability of a merozoite infecting a RBC of age. $a$ at time $t$ & dimensionless  \\ \hline
			$b_{i}(t)$& Rate at which adaptive immunity cell $i$ destroys iRBC. 			  &$1/(day \cdot cells/ul)$ 		\\ \hline
			$o_{i}(t)$& Rate at which innate immune cell enters circulation. &$cells/ul/day$			\\ \hline
			$l_{i}(t)$& Rate at which adaptive immune cell enters circulation. &$cells/ul/day$			\\ \hline

    \end{tabular}
		}
\end{center}
%--------------------------------------
\begin{center}
		{\scriptsize
    \begin{tabular}{ | l | p{5cm} | l | }
			\hline
			          			& Description & Unit 		 \\ \hline
			\multicolumn{3} {|c|}{Parameters} 
			\\ \hline
			$V$				  & Speed at which parasite ages, equal to 1/$\alpha$.  & $1/(day)$  \\ \hline
			$\varphi_{0}$   & Normal circulating RBC concentration, equal to $\varphi(0). $   & $cells/ul$			\\ \hline
			$\gamma$  			& Average number of merozoite produced by a single iRBC. 			  &$dimensionless$	   \\ \hline
			$\theta_{i}$    & Rate at which innate immunity cell $i$ destroys RBC. 		  & $1/(day\cdot cells/ul)$	 \\ \hline				
			$\psi_{i}$      & Rate at which adaptive immunity cell $i$ destroys RBC. 	 			  & $1/(day \cdot cells/ul)$ \\ \hline    
			$\phi_{i}$			& Rate at which innate immunity cell $i$ destroys iRBC. 			  & $1/(day \cdot cells/ul)$ \\ \hline	
			$\beta_{i}$			& Rate at which innate immunity cell decays.  & $1/day$	  \\ \hline
			$\delta_{i}$		& Rate at which adaptive immunity cell decays.  & $1/day$	 	  \\ \hline
			$\tau_{i}$			& Rate at which innate immunity cell decay due to contact with iRBC. & $1/(day \cdot cells/ul)$\\ \hline
			$\vartheta_{i}$			& Rate at which adaptive immunity cell decay due to contact with iRBC. & $1/(day \cdot cells/ul)$   \\ \hline
			$\lambda_{i}$		& Coefficient for change of adaptive immunity effector strength. & $1/cells/ul$	  \\ \hline
			$\varpi_{i}$		& Normal rate of production of innate immune cell $i$. 			  & $cells/ul/day$    \\ \hline
			$\sigma_{i}$		& Normal rate of production of adaptive immune cell $i$.  			  & $cells/ul/day$ \\ \hline
			$\epsilon_{i}$	& Maximum rate of production of innate immune cell $i$. 			  & $cells/ul/day$  \\ \hline
			$\varrho_{i}$		& Maximum rate of production of adaptive immune cell $i$. 			  &  $cells/ul/day$	 \\ \hline
			$\nu_{i}$				& Maximum adaptive immunity effector $i$ strength. 			  & $1/day/cells/ul$ \\ \hline
			$\eta_{i}$			& Coefficient for change of production of innate immune cell $i$.  			  & $1/cells/ul$ \\ \hline
			$\omega_{i}$		& Coefficient for change of production of adaptive immune cell $i$.  			  & $1/cells/ul$ \\ \hline
			$M_{i}$         &	Amount of parasite where the production of innate immune cell $i$. increase the most	& $cells/ul$\\ \hline
			$R_{i}$         & Amount of parasite where the production of adaptive immune cell $i$. increase the most	&	$cells/ul$\\ \hline
			$xi_{i}$				&	Amount of parasite where the production of adaptive immune cell $i$. effector strength increase the most		& $cells/ul$		\\ \hline
			$\varepsilon$		& Coefficient for change of erythropoiesis. 			  & $1/cells/ul$	  \\ \hline
			$\varsigma$				& Number of RBC entering blood stream. 			  & $cells/ul/day$	    \\ \hline
			$T_{d}$					& Delay of hematopoietic response. 							& $day$	\\ \hline

    \end{tabular}
		}
\end{center}
%--------------------------------------

%\section*{Acknowledgments}
%This project was funded in part by Federal funds from the US National Institute of Allergy and Infectious Diseases, National Institutes of Health, Department of Health and Human Services under contract \#HHSN272201200031C (PI: Mary Galinski), which supports the Malaria Host-Pathogen Interaction Center (MaHPIC).

%\bibliographystyle{abbrv}
%\bibliography{Thesis}	

\hfill 

\end{document}